\documentclass[twocolumn,showpacs,preprintnumbers,amsmath,amssymb]{revtex4-2} 
\usepackage{graphicx}
\usepackage{dcolumn}
\usepackage{bm}
\usepackage{tabularx}
\usepackage{ulem} 
\newcolumntype{M}{>{\centering\arraybackslash}m{1.85cm}}
\usepackage[export]{adjustbox}
\usepackage{float}
\usepackage{braket}
\usepackage{lipsum}
\usepackage{amsmath}
\graphicspath{{figure/}}
\usepackage{xcolor}   

\usepackage{caption}
\usepackage{subcaption}

\usepackage{hyperref}
\hypersetup{
    colorlinks=true, 
    linktoc=all,     
    urlcolor  = cyan,
    citecolor = blue,
    linkcolor=blue,  
}

\makeatletter
\newcommand{\colorcaption}[2][]{%
  \begingroup%
  \renewcommand{\@caption@fignum@sep}{ (Color online). }%
  \caption[#1]{#2}%
  \endgroup%
}
\makeatother
\newcommand{\orcid}[1]{\href{https://orcid.org/#1}{\hskip2pt\includegraphics[width=9pt]{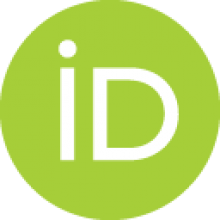}}}

\begin{document}

\title{ Isoscalar, isovector and orbital contributions in $M1$ transitions from analogous $M1$ and Gamow-Teller transitions in $T=\frac{1}{2}$ mirror nuclei }

\author{Subhrajit Sahoo\orcid{0000-0001-8000-2150}}
\email{s$\_$sahoo@ph.iitr.ac.in}
\affiliation{Department of Physics, Indian Institute of Technology Roorkee, Roorkee 247667, India}

\author{Praveen C. Srivastava\orcid{0000-0001-8719-1548}}
\email{ praveen.srivastava@ph.iitr.ac.in}
\affiliation{Department of Physics, Indian Institute of Technology Roorkee, Roorkee 247667, India}

\date{\hfill \today}

\begin{abstract}
The isoscalar and isovector components and their contributions to $M1$ transitions are discussed in the odd-$A$, $T=1/2$ mirror nuclei with mass number ranging from $A=23$ to 37. The orbital contributions in various $M1$ transitions and ground state magnetic moments are calculated by comparing analogous $M1$ and Gamow-Teller transitions between mirror pairs. The orbital contributions in different $M1$ transitions are explained on the basis of configurations of the initial and final states involved. In magnetic moments, the orbital contributions are found to be dependent on the deformation and single-particle nature of the states. All the $T=1/2$ mirror pairs are studied using isospin non-conserving interaction. The results are also compared with predictions from \textit{ab initio} effective interaction derived from realistic nuclear forces.
\end{abstract}


\maketitle
\section{Introduction}
The magnetic dipole ($M1$) and weak Gamow-Teller (GT) transitions provide crucial information for understanding the structure and configurations of atomic nuclei. The GT transition is induced by the spin charge exchange ($\sigma \tau$) operator, while the $M1$ transition involves both orbital ($l$) and spin ($\sigma$) angular momentum operators caused by the electromagnetic interaction \cite{BohrMottelsonBook}. Considering the charge-independent nature of strong interaction or isospin symmetry, the $M1$ operator can be decomposed into isoscalar ($l$ and  $\sigma$) and isovector ($l \tau$ and  $\sigma \tau$) components where the isovector (IV) or particularly $\sigma \tau$ component is dominant \cite{27AlSi}. So, the $M1$ and GT operators have similar characteristics, although they have different meson-exchange current contributions \cite{Brown1987, Etchegoyen1988, Sagawa2018, Matsubara2021}. The studies of similarities/differences between the $M1$ and GT transitions between analogous states and understanding their origins are essential for gaining deeper insights into the microscopic nature of $M1$ transitions.

The mirror nuclei offer unique opportunities to investigate and analyze different components of $M1$ transition. In the low-energy regime, the structural properties of mirror nuclei are equivalent, and they exhibit analogous states at similar excitation energies \cite{Wendt2014, AIPConf2005, Lenzi2001}. Extensive investigations into $M1$ and GT transitions between analogous states have been conducted in various mirror nuclei or isospin multiplets through different experimental methods, such as charge exchange reactions \cite{FujitaReview, Fujita_A27, 23NaMg, 25MgAl, Fujita_A26, Fujita_A25, Shimbara_A37, Adachi2012}. The contributions of the isovector orbital term ($l \tau$) to $M1$ transitions, which can be obtained by comparing analogous $M1$ and GT transitions, have been in focus. The $l \tau$ term can interfere constructively or destructively with the $\sigma \tau$ term and is responsible for enhancement or reduction in $B(M1)$ strength compared to $B(GT)$ strength. The experimental and theoretical analyses performed on some $T=1/2$ mirror nuclei 
in Refs. \cite{27AlSi,23NaMg, 25MgAl} reveal that the orbital contributions are largely dependent on the configurations involved in the transitions and the structure of nuclei. Large orbital contributions are observed in deformed nuclei.

In this article, we aim to study the analogous $M1$ and GT transitions between low-energy states in the $T_z= \pm 1/2$ odd-mass mirror pairs with $A=23$ to 37 within $sd$ model space. The isoscalar (IS) and isovector (IV) components are calculated for various $M1$ transitions between low-lying states, and the corresponding orbital contributions to the $M1$ transitions have been analyzed in detail, followed by a similar discussion of ground state (g.s.) magnetic moments. They are compared with available measured data and predictions are made where experimental data are not available yet. The transitions involving large orbital contributions and their configurations have been investigated in all the $T=1/2$ mirror pairs.

The interacting shell model, which serves as one of the most powerful tools to provide a unified description of both the structure and decays of atomic nuclei, has been used for this purpose. However, it is worth noting that most shell model interactions are founded on isospin symmetry. But, the Coulomb interaction between protons and other isospin-breaking effects of strong interaction can change the energy of analogous states in mirror nuclei and can cause asymmetry in the $M1$ decays \cite{Lam2013, Kaneko2017}. Hence, we have addressed the mirror pairs with the recently developed phenomenological $sd$ shell interaction USDC \cite{USDCint} that accounts for all such effects. Further, the recent advances and development in the \textit{ab initio} methods have enabled the decoupling of effective interactions for valence space starting from the underlying $NN$ and $NNN$ forces. The valence space effective interactions derived using \textit{ab initio} approaches such as no core shell model (NCSM) \cite{ncsm_effint_sdshell}, in-medium similarity renormalization group (IM-SRG) \cite{imsrg_ensembleReference, imsrg_Miyagi}, and coupled cluster theory (CC) \cite{CCEIint} have shown promising results in explaining ground state, low-lying excitation energies, electromagnetic observables, and other collective properties of open-shell nuclei \cite{Priyanka2023, Na_work_NPA, Archana2017}. These effective interactions preserve the fundamental characteristics of strong nuclear forces such as charge symmetry, Coulomb interaction, and charge-dependent components of nuclear forces. In this work, we have also discussed the $M1$ and GT transitions and their respective components from valence space effective interaction derived using the coupled cluster approach, and have provided the \textit{ab initio} predictions for the $T=1/2$ mirror nuclei.

The paper is organized in the following way. The formalism and important expressions related to the evaluation of different components of $M1$ transitions are briefly described in Sec. \ref{Sec2}. The results for IS and IV components and orbital contributions in various $M1$ transitions and ground state magnetic moments of $T=1/2$ mirror pairs are presented in Sec. \ref{Sec3}. The summary and conclusions are given in Sec. \ref{Sec4}.

\section{Formalism} \label{Sec2}
To obtain the isoscalar and isovector components, and other parameters of $M1$ transitions and magnetic moments,
we have explicitly followed the formalism as described by Fujita \textit{et al.} in Refs. \cite{27AlSi,23NaMg,25MgAl}.  In this section, we briefly outline the important expressions for these quantities.

The $M1$ transition or magnetic moment operator $\Vec{\mu}$ consisting of an orbital part $g_l\Vec{l}$ and spin part $(1/2)g_s\Vec{\sigma}$ can be written in terms of IS and IV components as 
\begin{equation} \label{eq1}
    \Vec{\mu} = \sum_{i=1}^{A} \left[  ( g_l^{IS}\Vec{l}_i+(1/2)g_s^{IS}\Vec{\sigma}_i ) -  ( g_l^{IV}\Vec{l}_i+(1/2)g_s^{IV}\Vec{\sigma}_i )\tau_{zi}     \right]   
\end{equation}
in the units of nuclear magneton ($\mu_N$). Here $\tau_{zi}=-1$ for protons and +1 for neutrons. The IS and IV $g$- factor coefficients are give by $g_l^{IS}=\frac{1}{2}(g_l^{\pi}+g_l^{\nu})$, $g_s^{IS}=\frac{1}{2}(g_s^{\pi}+g_s^{\nu})$, $g_l^{IV}=\frac{1}{2}(g_l^{\pi}-g_l^{\nu})$, and $g_s^{IV}=\frac{1}{2}(g_s^{\pi}-g_s^{\nu})$, where $g_l$ and $g_s$ correspond to the orbital and spin $g$ factors of bare protons ($\pi$) and neutrons ($\nu$), respectively.

The $M1$ transition strength [$B(M1)$] between initial state with spin $J_i$ and isospin $T_i$ and final state with corresponding quantum numbers $J_f$ and $T_f$, respectively is given by
\begin{equation}\label{eq2}
    B(M1) = \frac{1}{2J_i+1} \frac{3}{4\pi} {\vert \langle J_fT_fT_{zf} \Vert \Vec{\mu} \Vert J_iT_iT_{zi} \rangle \vert}^2,
\end{equation}
and the magnetic moment of a state with spin $J$ and isospin $T$ is given by 
\begin{equation}\label{eq3}
    \mu = \sqrt{\frac{J}{(J+1)(2J+1)}} \langle JTT_z \Vert \Vec{\mu} \Vert JTT_z \rangle .
\end{equation}
By applying the Wigner-Eckart theorem in the isospin space, the $B(M1)$ operator can be written as \cite{larry}
\begin{equation}\label{eq4}
    B(M1) =\frac{1}{2J_i+1} \frac{3}{4\pi} {\left[ M^{IS}_{M1} - \frac{C_{M1}}{\sqrt{2T_f+1}}M_{M1}^{IV} \right]}^2,
\end{equation}
where $C_{M1}=(T_iT_{zi}10\vert T_f T_{zf})$ is the Clebsch-Gordan (or C.G.) coefficient. The IS and IV $M1$ matrix elements of Eq. \eqref{eq4} are given by
\begin{equation}\label{eq5}
    M^{IS}_{M1} = g_l^{IS}M_{M1}(l) + \frac{1}{2} g_s^{IS}M_{M1}(\sigma)
\end{equation}
and 
\begin{equation}\label{eq6}
    M^{IV}_{M1} = g_l^{IV}M_{M1}(l \tau) + \frac{1}{2} g_s^{IV}M_{M1}( \sigma \tau).
\end{equation}
Here $M_{M1}(x)$ represent the reduced matrix elements defined by $\langle J_fT_f \Vert \sum_{i=1}^A \Vec{x}_i \Vert J_iT_i \rangle$, where $\Vec{x}=\lbrace{ l,\sigma,l \tau, \sigma \tau \rbrace}$. The IV spin term and hence the $M^{IV}_{M1}$ matrix element of $M1$ transition is usually larger than $M^{IS}_{M1}$ due to the coefficient $g_s^{IV}$. Therefore, in the $M1$ transition, the IS term may interfere constructively or destructively with the IV term. Additionally, when the IS term is very small, the orbital IV term [$g_l^{IV}M_{M1}(l \tau)$] interferes constructively or destructively with the IV spin term [$\frac{1}{2} g_s^{IV}M_{M1}( \sigma \tau)$]. The contributions of the IS term or the orbital IV term to an $M1$ transition are discussed later in this section.

A similar application of the Wigner-Eckart theorem to the magnetic moment operator results in
\begin{equation}\label{eq7}
    \mu = \sqrt{\frac{J}{(J+1)(2J+1)}} \left[ M^{IS}_{M1} - \frac{C_{M1}}{\sqrt{2T+1}}M_{M1}^{IV} \right],
\end{equation}
where $C_{M1}=(TT_z10\vert T T_{z})$.

For analogous $M1$ transitions between $T_z= \pm T$ mirror nuclei, under the assumption that the IS part is smaller than the IV part, the IS and IV parts of the $M1$ transitions are derived from Eq. \eqref{eq4} as (see Ref. \cite{27AlSi})
\begin{equation}\label{eq8}
    B_{IS}(M1) = \frac{1}{4} { \left[ \sqrt{B(M1)}_{T_z=+T} - \sqrt{B(M1)}_{T_z=-T} \right] }^2,
\end{equation}
and
\begin{equation}\label{eq9}
    B_{IV}(M1) = \frac{1}{4} { \left[ \sqrt{B(M1)}_{T_z=+T} + \sqrt{B(M1)}_{ T_z=-T} \right] }^2.
\end{equation}
Similarly, using Eq. \eqref{eq7}, the IS and IV components of the magnetic moment can be written as $\mu_{IS}= \frac{1}{2}( \mu_{T_z=T} + \mu_{T_z=-T})$ and $\mu_{IV}= \frac{1}{2}( \mu_{T_z=T} - \mu_{T_z=-T})$.

The IS contribution to the $M1$ transitions can be analyzed through a parameter $R_{IS}$ as
\begin{equation}\label{eq10}
    R_{IS} = \frac{B(M1)}{B_{IV}(M1)}.
\end{equation}
The value of $R_{IS}$ larger than unity indicates that the IS term constructively interferes with the IV term and vice-versa. The constructive or destructive contributions of the IS part are reversed in the $T_{z}=\pm 1/2$ pair, and a seesaw-like relationship of $R_{IS}$ is expected in the $T_{z}= 1/2$ and $T_{z}=- 1/2$ nuclei. For the magnetic moment, the $R_{IS}$ parameter can be written as 
\begin{equation}\label{eq11}
    R_{IS} = { \left( \frac{\mu}{\mu_{IV}} \right) }^2.
\end{equation}

The Gamow-Teller (GT) transition strength between initial (i) and final (f) states with $J$ and $T$ as good quantum numbers is given by
\begin{equation} \label{eq12}
    B(GT) = \frac{1}{2}  \frac{1}{2J_i+1}  \frac{C_{GT}^2}{2T_f+1} {\left[  M_{GT}(\sigma \tau) \right]}^2,
\end{equation}
where $M_{GT}(\sigma \tau)=\langle J_fT_f \Vert \sum_{i=1}^A  \Vec{\sigma}_i \Vec{\tau}_i \Vert  J_iT_i \rangle$ and $C_{GT}=(T_iT_{zi} 1\pm1 \vert T_fT_{zf})$ is the C.G. coefficient.
For analogous states, the IV spin operator ($\sigma \tau$) is similar in both $M1$ and GT transition. However, the contributions of meson exchange currents (or MEC) are different for $M1$ and GT operators. So, for $M1$ and GT transitions between analogous states, a relation can be established between the IV spin type GT matrix [$M_{GT}(\sigma \tau)$] and the IV spin term of the $M1$ transition [$M_{M1}(\tau \sigma)$] through a parameter $R_{MEC}$ defined as 
\begin{equation}\label{eq13}
    R_{MEC}=\frac{{[M_{M1}(\sigma \tau) ]}^2}{{[M_{GT}(\sigma \tau) ]}^2}.
\end{equation}
The MEC contributions to $M1$ and GT operators, including core polarization effects, are taken care of by the parameter $R_{MEC}$ (for details, see \cite{27AlSi}). This implies, when $M_{M1}( \sigma \tau)$ is dominant in the $M1$ transition, the $B(M1)$ value in one of the mirror pair should be $B(M1) \approx \frac{3}{8\pi} {(g_s^{IV})}^2 \frac{C_{M1} ^2}{C_{GT} ^2}R_{MEC}B(GT)$ \cite{23NaMg}. Any significant deviation from this value will imply constructive or destructive interference of the orbital term with the spin term in the $M^{IV}_{M1}$ element. The orbital contribution to the IV term of  the $M1$ transition can be readily identified by the ratio \cite{27AlSi}

\begin{equation}\label{eq14}
    R_{oc} = \frac{8\pi}{3{(g_s^{IV})}^2}\frac{1}{R_{MEC}} \frac{C^2_{GT}}{C^2_{M1}} \frac{B_{IV}(M1)}{B(GT)},
\end{equation}
and for the magnetic moment $R_{oc}$ becomes
\begin{equation} \label{eq15}
    R_{oc} = \frac{2(J+1)}{J{(g_s^{IV})}^2} \frac{C^2_{GT}}{C^2_{M1}} \frac{1}{R_{MEC}}\frac{\mu^2_{IV}}{B(GT)},
\end{equation}
where the ratio of squared of C.G. coefficients is 2 for $T=1/2$ nuclei. Here, $R_{oc} > 1$ indicates that the orbital part makes constructive interference with the spin term in $M_{M1}^{IV}$ to enhance the $M1$ transition and vice versa unless the transition strength is very small, in which case the spin term is hindered and much smaller than the orbital one, i.e., $\vert g_l^{IV}M_{M1}(l \tau) \vert >> \vert \frac{1}{2} g_s^{IV}M_{M1}(\tau \sigma) \vert$.

To find the orbital contribution to the $M1$ operator or $R_{oc}$, one needs to calculate the IV component of the $B(M1)$ transition strength. However, in many cases, for analogous $M1$ transitions in mirror pairs, the $B(M1)$ strengths are not available or not yet measured in one of the nuclei within mirror pairs. 
In such cases, by approximating the IV term to be dominant, or $B(M1) \approx B_{IV}(M1)$, the combined contribution of IS and orbital terms to the IV spin term can be inferred through a parameter $R_{ISO}$ as follows \cite{23NaMg,25MgAl}:
\begin{equation} \label{eq16}
    R_{ISO} = \frac{8\pi}{3{(g_s^{IV})}^2} \frac{C^2_{GT}}{C^2_{M1}} \frac{1}{R_{MEC}}\frac{B(M1)}{B(GT)}.
\end{equation}
Since the IS contributions are usually very small, the $R_{ISO}$ values closely resemble with $R_{oc}$.  

In the experimental works \cite{23NaMg,27AlSi,25MgAl}, the average value of $R_{MEC}$ was taken as 1.25 for $sd$-shell nuclei, and the IS and IV components of the $M1$ transition or the magnetic moment, $R_{oc}$, and $R_{ISO}$ parameters were deduced. In order to facilitate the comparison between our calculated results and these experimental findings, we have adopted the same value for $R_{MEC}$ and $g_s^{IV}=4.706$ in our calculations. In particular, Eqs. \eqref{eq8} to \eqref{eq11} and Eqs. \eqref{eq14} to \eqref{eq16} are used to calculate various components of $M1$ transitions and different parameters, as described above.

In the framework of the shell model, the $M1$ and GT operators for the mirror pairs are calculated from the universal $sd$- shell variant or USD-based USDC interaction. Starting from a renormalized G-matrix approach \cite{Gmatrix}, the two-body matrix elements (TBMEs) of USD-based interactions are determined and adjusted to fit several experimental datasets \cite{USDA_USDB}. The USDC interaction incorporates the Coulomb interaction between protons and other isospin-breaking parts of the strong interaction \cite{USDCint}. The g.s., low-lying excited states, and other isospin-symmetry breaking effects are well captured by the USDC interaction in most of the $sd$- shell nuclei within experimental uncertainty compared to other existing interactions. The $M1$ and GT components are also compared to results from an \textit{ab initio} coupled-cluster effective interaction (CCEI) derived from realistic chiral $NN$ and $NNN$ forces using the coupled-cluster approach.
The CCEI is decoupled nonperturbatively from the chiral potentials \cite{N3LO, N2LO} for open-shell nuclei with valence protons and neutrons and retains the properties of the original nucleon-nucleon interaction\cite{CCEIint}. The CCEI interaction provides a good description of the ground state binding energies, low-lying excitation spectra as well as collective behaviors of open-shell nuclei from first principles. The shell model Hamiltonians are diagonalized using NushellX \cite{nushellx} and KSHELL \cite{kshell}.


\section{Results and Discussion} \label{Sec3}
In the shell-model approach, various factors, such as the effects of core polarization, model space truncation, MEC contributions, etc., are taken into account by introducing effective model space operators \cite{Brown1987, Etchegoyen1988}. Such studies were performed with USD-type Hamiltonians (USD, USDA, and USDB) in Ref. \cite{effE2M1GT}, and effective $E2$, $M1$, and GT operators or effective charges, $g$- factors ($g^{eff}$), and quenching factor ($q$), respectively were proposed for these Hamiltonians. It was found that the operators are less sensitive to the USD Hamiltonians, with their values remaining approximately the same in all USD-type Hamiltonians. Following this observation, we have used the same effective $g$- factors [${(g_l^{\pi})}^{eff}=1.159,{(g_l^{\nu})}^{eff}=-0.09$ and ${(g_s^{\pi})}^{eff}= 5.15,{(g_s^{\nu})}^{eff}=-3.55$] and $q=0.76$, as determined for the USDB Hamiltonian, to calculate the $M1$ and GT transition matrix elements from USD-based USDC interaction. The analysis of GT strengths performed in Ref. \cite{qfactorCCEI} using CCEI yields  $q=0.78$, approximately the same as that obtained for USDB. Since the major components of $M1$ and GT transitions are similar ( $\sigma \tau$ ), identical effective $g$- factors can be expected for CCEI as those found in the USDB Hamiltonian. Hence, to ensure consistency, we have employed the same effective $g$- and quenching factors for CCEI as those employed for USDC.

The mirror nuclei exhibit analogous states ($J^{\pi}$) with similar excitation energies ($E_x$). The $M1$ transition strengths between the ground state and lower excited states within the mirror nuclei and the GT transition strengths for the analogous transitions between the mirror pair are calculated for $T=1/2$ nuclei from $A=23$ to $37$ in Table \ref{tab_1}. The IS and IV components of the $M1$ transitions [$B_{IS}(M1)$ and $B_{IV}(M1)$], IS contributions in each mirror nuclei ($R_{IS}$) and orbital contribution to the IV term of $M1$ transitions ($R_{oc}$ or $R_{ISO}$) are listed in different columns of Table \ref{tab_1}. The first row ( or the transition from g.s. to g.s.) for each mirror pair represents the magnetic moment ($\mu$) instead of $B(M1) \uparrow$, thus indicating the IS ($\mu_{IS}$) and IV components ($\mu_{IV}$), along with IS and orbital contributions to the g.s. magnetic moments. The experimental uncertainties associated with the magnetic moments are much smaller than their respective values, and hence, they have not been included. In the next section, we will see that the  $B_{IS}(M1)$ values are very small. When the experimental values are very small, sometimes the errors associated with them may appear larger than their corresponding values. In those cases, the experimental errors in $B_{IS}(M1)$ are not considered (as followed in Ref. \cite{23NaMg}). The phenomenological interactions have supremacy in describing the experimental behaviors and are always used as a benchmark. So, we have made a one-to-one comparison of the experimental data wherever available, with the results obtained from the USDC interaction in Table \ref{tab_1}. Then, the predictions of \textit{ab initio} CCEI are compared with the USDC results through various figures. First, we discussed the $M1$ components in each mirror pair, then we analyze the g.s. magnetic moments and their corresponding components for all mirror pairs within the $sd$ shell.

\begin{table*}
\addtolength{\tabcolsep}{-0.15mm}
\centering
\caption{ 
Experimental and calculated data for $B(M1) \uparrow$, $B(GT)$, and their corresponding components in $T=\frac{1}{2}$ mirror nuclei. For more details, see text. }

 
\begin{tabular}{ lccccccccccccc } 
 \hline
 \hline
   & &  \multicolumn{2}{c}{$E_x$} & & \multicolumn{2}{c}{$B(M1) \uparrow$ } &  &  &   &\multicolumn{2}{c}{$R_{IS}$} &   \\
  \cline{3-4}
  \cline{6-7}
  \cline{11-12}

\hspace{-0.3cm}  $\mathbf{ ^{23}Na  \rightarrow ^{23}Mg }$\\ 
\textbf{Expt.} \cite{23NaMg} \\
$A$ & $J^{\pi}$ &\hspace{2mm} $^{23}$Na &\hspace{2mm} $^{23}$Mg & &\hspace{2mm} $^{23}$Na &\hspace{2mm} $^{23}$Mg &\hspace{2mm} $B(GT)$ &\hspace{2mm} $B_{IS}(M1)$ &\hspace{2mm} $B_{IV}(M1)$  &\hspace{2mm} $^{23}$Na &\hspace{2mm} $^{23}$Mg &\hspace{2mm} $R_{oc}$ &$R_{ISO}$ \\
 \hline
\\
23 & $3/2^+_1$  &  0.0  & 0.0 & &2.218 \footnotemark[1] & -0.536 \footnotemark[1] &0.190(4)
            & 0.841 \footnotemark[2]& 1.337 \footnotemark[3] & 2.59 & 0.15 & 2.3(1) &    \\
   & $5/2^+_1$  &  0.440  & 0.459 & &0.554(34)  & 0.591(64) &0.146(6)
            & $1 \times 10^{-4}$ & 0.57(4) & 0.97(7) & 1.03(13) & 2.4(2) &    \\
   & $1/2^+_1$  &  2.391  & 2.359 & &0.0017(3)  & 0.0017(4) &0.055(4)
            & $1 \times 10^{-8}$ & 0.0017(3) & 1.0(2) & 1.0(2) & 0.019(3) &    \\
    & $3/2^+_2$  &  2.982  & 2.906 & &0.292(41) &    & 0.193(11)
            &   &   &   &   &  &0.92(14)   \\
   & $5/2^+_2$  &  3.914  & 3.860 & &0.090(15)  &   &0.055(4)
            &   &   &   &   &  &0.99(18)   \\
   & $1/2^+_2$  &  4.430  & 4.357 & &1.02(7)  &   &0.250(13)
            &   &   &   &   &   &2.48(22)   \\
    & $5/2^+_3$  &  5.379  & 5.291 & &0.33(12)  &   &0.066(5)
            &   &   &   &   &  &3.0(11)   \\
\textbf{USDC} \\
23 & $3/2^+_1$  &  0.0  & 0.0 & &2.162 \footnotemark[1] & -0.396 \footnotemark[1] &0.164
            & 0.883 \footnotemark[2] & 1.279 \footnotemark[3] & 2.85 & 0.09  &  2.40 &    \\
   & $5/2^+_1$  &  0.408  & 0.397 & &0.537  & 0.457 &0.134
            & $8 \times 10^{-4}$ & 0.496 & 1.08 & 0.92 & 2.24 &    \\
   & $1/2^+_1$  &  2.190  & 2.162 & &0.020  & 0.002 &0.071
            & $2 \times 10^{-3}$ & 0.008 & 2.31 & 0.23 & 0.07 &    \\
    & $3/2^+_2$  &  2.751  & 2.746 & &0.333 & 0.230   & 0.207
            & $2 \times 10^{-3}$ & 0.280  & 1.19  & 0.82  & 0.82 &0.97   \\
   & $5/2^+_2$  &  3.748  & 3.739 & &0.058  & 0.038  &0.046
            & $5 \times 10^{-4}$  & 0.047  & 1.22  & 0.80  &  0.62 & 0.76 \\
   & $1/2^+_2$  &  4.414  & 4.425 & &1.047  & 0.863  &0.241
            & $2 \times 10^{-3}$  & 0.953  & 1.09  & 0.91  & 2.40 &2.63 \\
    & $5/2^+_3$  &  5.363  & 5.367 & &0.345  & 0.297  &0.125
            & $4 \times 10^{-4}$  & 0.321  & 1.07  & 0.93  & 1.55 &1.67 \\
\\

\hspace{-0.3cm} $ \mathbf{^{25}Mg  \rightarrow ^{25}Al}$\\
\textbf{Expt.} \cite{25MgAl}\\
$A$ & $J^{\pi}$ &\hspace{2mm} $^{25}$Mg &\hspace{2mm} $^{25}$Al & &\hspace{2mm} $^{25}$Mg &\hspace{2mm} $^{25}$Al &\hspace{2mm} $B(GT)$ &\hspace{2mm} $B_{IS}(M1)$ &\hspace{2mm} $B_{IV}(M1)$  &\hspace{2mm} $^{25}$Mg &\hspace{2mm} $^{25}$Al &\hspace{2mm} $R_{oc}$ &$R_{ISO}$ \\
 \hline
\\
25 & $5/2^+_1$  &  0.0  & 0.0 & & -0.855 \footnotemark[1] & 3.645  \footnotemark[1] &0.408(2)
            & 1.395 \footnotemark[2] & -2.250 \footnotemark[3] & 0.14 & 2.62  &  2.51   \\
   & $3/2^+_1$  &  0.975  & 0.945 & &0.0011(1)  &   &0.003(1)
             &   &   &   &   &      \\
   & $7/2^+_1$  &  1.612  & 1.613 & &0.83(12)  &   &0.165(7)
             &   &   &   &   &  &3.1(8)   \\
   & $5/2^+_2$  &  1.790  & 1.965  &  & 0.0014(2) &   &0.019(2)
             &   &   &   &   &      \\
   & $3/2^+_2$  &  2.674  & 2.801 & &0.007(2)  &   &0.017(2)
            &   &   &   &   &      \\
\\
\textbf{USDC}\\
25 & $5/2^+_1$  &  0.0  & 0.0 & &-0.861 \footnotemark[1]& 3.757 \footnotemark[1] &0.360
            & 1.448 \footnotemark[2] & -2.309 \footnotemark[3] & 0.14  & 2.65 & 3.0 &   \\
   & $3/2^+_1$  & 1.072  & 1.015 & &0.003  & 0.005  &0.003
             & $6 \times 10^{-5}$ & 0.004  & 0.76  & 1.27 &  0.80 &    \\
             
   & $7/2^+_1$  &  1.708  & 1.723 & &0.810  & 0.881  &0.131
             & $4 \times 10^{-4}$ & 0.845 & 0.95  & 1.04  & 3.90 &3.70 \\
             
   & $5/2^+_2$  & 2.012  & 1.882  &  & 0.003 & 0.002  &$7 \times 10^{-5}$
             & $2 \times 10^{-5}$ & 0.002 & 1.21  & 0.81 &  21.40 &    \\            
   & $3/2^+_2$  &  2.834  & 2.739 & &0.014  & 0.019  &0.010
            & $1 \times 10^{-4}$ & 0.016  & 0.85  & 1.15 & 0.99 &    \\
\\

\hspace{-0.3cm} $\mathbf{ ^{27}Al  \rightarrow ^{27}Si } $\\
\textbf{Expt.} \cite{27AlSi}\\
$A$ & $J^{\pi}$ &\hspace{2mm} $^{27}$Al &\hspace{2mm} $^{27}$Si & &\hspace{2mm} $^{27}$Al &\hspace{2mm} $^{27}$Si &\hspace{2mm} $B(GT)$ &\hspace{2mm} $B_{IS}(M1)$ &\hspace{2mm} $B_{IV}(M1)$  &\hspace{2mm} $^{27}$Al &\hspace{2mm} $^{27}$Si &\hspace{2mm} $R_{oc}$ & $R_{ISO}$\\
 \hline
\\
27 & $5/2^+_1$  &  0.0  & 0.0 &  & 3.642 \footnotemark[1]& -0.865 \footnotemark[1]&  0.307(44)
            & 1.388 \footnotemark[2] & 2.254 \footnotemark[3] & 2.6 & 0.15  & 3.3(5) &     \\
   & $3/2^+_1$  &  1.014  & 0.957 & &0.015(1)  & 0.019(1) & 2.2(3)
   $\times 10^{-4}$ & 6(5)$\times 10^{-5}$  & 0.017(1)  & 0.89(7)  & 1.12(9)  &  51.3(80) &    \\
   & $7/2^+_1$  &  2.211  & 2.164 & &0.150(4)  & 0.102(10)  &0.079(6)
             & 1.2(5)$\times 10^{-3}$ & 0.125(6) & 1.20(6) & 0.82(8)  & 0.96(12) &      \\
 
   & $5/2^+_2$  &  2.735  & 2.648 &  & 0.046(7)& 0.022(5) &  0.039(4)
            & 1.1(8)$\times 10^{-3}$  & 0.033(4)  & 1.40(28) & 0.67(18) &  0.51(10) &    \\
    & $3/2^+_2$  &  2.982  & 2.866 &  & 0.245(13)&   &  0.173(12)
            &   &   &   &   &  &0.85(3)     \\
    & $3/2^+_3$  &  3.957  & 3.804 &  & 0.145(12)&   &  0.079(7)
            &   &   &   &   &  &1.11(1)   \\
    & $5/2^+_3$  &  4.410  & 4.289 &  & 0.226(28)& 0.075(37) & 0.097(9)
            & 0.010(7)  &  0.140(28) & 1.61(37) & 0.53(28)  &  0.88(19) &    \\

 \\
\textbf{USDC}\\ 
27 & $5/2^+_1$  &  0.0  & 0.0 &  & 3.576 \footnotemark[1] & -0.710 \footnotemark[1]&  0.275
             & 1.433 \footnotemark[2] & 2.143 \footnotemark[3]& 2.78  &  0.11 & 3.38 &    \\
   & $3/2^+_1$  &  1.063  & 1.117 & &0.001  & 0.007 & 0.003 
             & $7 \times 10^{-4}$  & 0.003  & 0.30  & 2.10  & 0.67 &     \\
   & $7/2^+_1$  &  2.289  & 2.346 & &0.128  & 0.100  &0.075
             & $4 \times 10^{-4}$  & 0.113  & 1.13  & 0.88  &  0.92 &    \\
    & $5/2^+_2$  &  2.692  & 2.733 &  & 0.074& 0.050 &  0.058
            & $6 \times 10^{-4}$  & 0.061  & 1.20  & 0.82  &  0.64    \\
    & $3/2^+_2$  &  2.843  & 2.835 &  & 0.347& 0.242  &  0.206
            & $2 \times 10^{-3}$  & 0.292  & 1.19  & 0.83  &  0.86 &1.02 \\
    & $3/2^+_3$  &  3.983  & 3.951 &  & 0.204& 0.110  &  0.067
            & $3 \times 10^{-3}$  & 0.153  & 1.33  & 0.72  & 1.38 & 1.84 \\
    & $5/2^+_3$  &  4.367  & 4.432 &  & 0.300& 0.220 & 0.090
            & $1 \times 10^{-3}$  & 0.258  & 1.16  & 0.85  &  1.74 &    \\
    & $7/2^+_2$  &  4.675  & 4.689 &  & 0.064& 0.037 &  0.019
            & $9 \times 10^{-4}$  & 0.050  & 1.30  & 0.74  &  1.58 &   \\

 \\
 \hline 

 \end{tabular}
					
	\end{table*}
\addtocounter{table}{-1}

\begin{table*}
\addtolength{\tabcolsep}{+0.3mm}
	\leavevmode
	 \centering  
	 \caption{ Continued.}\label{tab_1} 
	 \begin{tabular}{ lccccccccccccc } 
\hline
 \hline
   & &  \multicolumn{2}{c}{$E_x$} & & \multicolumn{2}{c}{$B(M1) \uparrow$ } &  &  &   &\multicolumn{2}{c}{$R_{IS}$} &   \\
  \cline{3-4}
  \cline{6-7}
  \cline{11-12}

\hspace{-0.3cm} $\mathbf{^{29}Si  \rightarrow ^{29}P} $\\
\textbf{Expt.} \cite{NNDC}\\
$A$ & $J^{\pi}$ &\hspace{2mm} $^{29}$Si &\hspace{2mm} $^{29}$P & &\hspace{2mm} $^{29}$Si &\hspace{2mm} $^{29}$P &\hspace{2mm} $B(GT)$ &\hspace{2mm} $B_{IS}(M1)$ &\hspace{2mm} $B_{IV}(M1)$  &\hspace{2mm} $^{29}$Si &\hspace{2mm} $^{29}$P &\hspace{2mm} $R_{oc}$ \\
 \hline
\\
29 & $1/2^+_1$  &  0.0  & 0.0 &  & -0.556 \footnotemark[1] & 1.235 \footnotemark[1] &   
            & 0.340 \footnotemark[2] & -0.895 \footnotemark[3] & 0.38  & 1.90  &      \\
   & $3/2^+_1$  & 1.273 & 1.383 & & 0.127(9) & 0.193(50) &   
            & $2 \times 10^{-3}$  & 0.158(23)  & 0.80(13)  & 1.22(36) &      \\
   & $3/2^+_2$  & 2.425 & 2.422 & & 0.233(28) & 0.233(78)  & 
             &   &   &   &   &      \\
    & $1/2^+_2$ & 4.840 & 4.759 &  & $>$0.086  &   &   
            &   &   &   &   &      \\
 \\

\textbf{USDC}\\
$A$ & $J^{\pi}$ &\hspace{2mm} $^{29}$Si &\hspace{2mm} $^{29}$P & &\hspace{2mm} $^{29}$Si &\hspace{2mm} $^{29}$P &\hspace{2mm} $B(GT)$ &\hspace{2mm} $B_{IS}(M1)$ &\hspace{2mm} $B_{IV}(M1)$  &\hspace{2mm} $^{29}$Si &\hspace{2mm} $^{29}$P &\hspace{2mm} $R_{oc}$ \\
 \hline

29 & $1/2^+_1$  &  0.0  & 0.0 &  & -0.500 \footnotemark[1] & 1.144 \footnotemark[1]&  0.141 
            & 0.322 \footnotemark[2] &  -0.822 \footnotemark[3] &  0.37 &  1.94 &  2.07   \\
   & $3/2^+_1$  & 1.263 & 1.337 & & 0.031 & 0.051 & 0.026  
            & $6 \times 10^{-4}$ & 0.040 & 0.77 & 1.26 & 0.94    \\
   & $3/2^+_2$  &  2.505 & 2.579 & & 0.422 & 0.585  & 0.322
             & $3 \times 10^{-3}$  & 0.500  & 0.84  & 1.17  &  0.94 \\
    & $1/2^+_2$ & 4.760  & 4.756 &  & 0.268 & 0.437 &  0.242 
            & $5 \times 10^{-3}$  & 0.347 & 0.77 & 1.25  &  0.87    \\

    & $3/2^+_3$  &  6.075 & 6.129 & & 0.004 & $1 \times 10^{-4}$  & 0.003
             & $7 \times 10^{-4}$  & 0.001  & 2.98  & 0.07  &  0.27 \\
    & $1/2^+_3$ & 6.656  & 6.717 &  & 0.154 & 0.283 &  0.112 
            & $5 \times 10^{-3}$  & 0.213 & 0.72 & 1.32  &  1.15    \\
\\

\hspace{-0.3cm} $\mathbf{^{31}P  \rightarrow ^{31}S} $\\
\textbf{Expt.} \cite{NNDC}\\
$A$ & $J^{\pi}$ &\hspace{2mm} $^{31}$P &\hspace{2mm} $^{31}$S & &\hspace{2mm} $^{31}$P &\hspace{2mm} $^{31}$S &\hspace{2mm} $B(GT)$ &\hspace{2mm} $B_{IS}(M1)$ &\hspace{2mm} $B_{IV}(M1)$  &\hspace{2mm} $^{31}$P &\hspace{2mm} $^{31}$S &\hspace{2mm} $R_{oc}$ \\
 \hline
\\
31 & $1/2^+_1$  &  0.0  & 0.0 &  & 1.131 \footnotemark[1] & (-)0.488 \footnotemark[1] &   
            & 0.325 \footnotemark[2] & 0.810 \footnotemark[3] & 1.95  & 0.36  &      \\
   & $3/2^+_1$  &  1.266  & 1.249 & &0.070(4)  & 0.072(18) &   
            & $3 \times 10^{-6}$  & 0.071(10)  & 0.98(14)  & 1.01(28)  &      \\
   & $1/2^+_2$  &  3.134  & 3.079 & &0.175(78)  &   & 
             &   &   &   &   &      \\
    & $3/2^+_2$ &  3.506  & 3.436 &  & 0.104(16) &   &   
            &   &   &   &   &      \\
 \\

\textbf{USDC}\\
31 & $1/2^+_1$  &  0.0  & 0.0 &  & 1.047 \footnotemark[1]& -0.427 \footnotemark[1] &  0.181 
            & 0.310 \footnotemark[2] &  0.737 \footnotemark[3] & 2.02  & 0.33  & 1.30    \\
   & $3/2^+_1$  &  1.179  & 1.186 & &0.019  & 0.007 & 0.037
            & $7 \times 10^{-4}$ & 0.012 & 1.55  & 0.57  &  0.20   \\
   & $1/2^+_2$  &  3.268  & 3.272 & &0.216  & 0.164  & 0.041
             & $9 \times 10^{-4}$ & 0.190  & 1.14 & 0.87 & 2.80    \\
   & $3/2^+_2$ &  3.605  & 3.644 &  & 0.195 & 0.113 &  0.153 
            & $3 \times 10^{-3}$  & 0.151  & 1.29  & 0.75  & 0.60   \\
   & $3/2^+_3$ &  4.392  & 4.412 & & 0.017  & 0.014  &  0.015 
            &  $3 \times 10^{-5}$  & 0.015   & 1.10   &  0.90  &  0.62   \\
   & $1/2^+_3$  &  5.013  & 5.020 & & 0.112 & 0.128    &  0.046
             & $1 \times 10^{-4}$  & 0.120   & 0.93  &  1.06 &   1.57   \\

 \\

\hspace{-0.3cm} $\mathbf{^{33}S  \rightarrow ^{33}Cl} $\\
\textbf{Expt.} \cite{NNDC}\\
$A$ & $J^{\pi}$ &\hspace{2mm} $^{33}$S &\hspace{2mm} $^{33}$Cl & &\hspace{2mm} $^{33}$S &\hspace{2mm} $^{33}$Cl &\hspace{2mm} $B(GT)$ &\hspace{2mm} $B_{IS}(M1)$ &\hspace{2mm} $B_{IV}(M1)$  &\hspace{2mm} $^{33}$S &\hspace{2mm} $^{33}$Cl &\hspace{2mm} $R_{oc}$ \\
 \hline
\\
33 & $3/2^+_1$  &  0.0  & 0.0 &  & 0.644 \footnotemark[1]& 0.755 \footnotemark[1]&   
            & 0.700 \footnotemark[2] &  -0.055 \footnotemark[3]& 134.64  & 185.05  &      \\
   & $1/2^+_1$  &  0.841  & 0.810 & &0.027(4)  & 0.015(5) &   
            & $4 \times 10^{-4}$  & 0.021(3)  &  1.31(30) & 0.73(27)  &      \\
   & $5/2^+_1$  &  1.967  & 1.986 & &0.057(21)  & 0.108(48)  & 
             & $2 \times 10^{-3}$  & 0.080(24)  & 0.71(33)  & 1.34(72)  &      \\
   & $3/2^+_2$ &  2.314  & 2.352 &  & 0.0077(14) & 0.004(2) &   
            & $1.5 \times 10^{-4}$  & 0.005(1)  & 1.35(40)  & 0.70(38)  &      \\
   & $5/2^+_2$ &  2.867  & 2.839 &  & 0.214(162) & 0.465(120) &   
            & $1.2 \times 10^{-3}$  & 0.327(112)  & 0.65(54)  & 1.42(61)  &    \\

 \\
\textbf{USDC}\\
33 & $3/2^+_1$  &  0.0  & 0.0 &  & 0.664 \footnotemark[1] & 0.841 \footnotemark[1]&  0.099 
            & 0.752 \footnotemark[2]& -0.088 \footnotemark[3] & 56.30 & 90.30 &  0.02    \\
   & $1/2^+_1$  &  0.893  & 0.899 & &0.011  & 0.021 &  0.007 
            & $4 \times 10^{-4}$  & 0.015  & 0.71  & 1.35  & 1.35    \\
   & $5/2^+_1$  &  1.956  & 1.965 & &0.016  & 0.031  & 0.021
             & $6 \times 10^{-4}$  & 0.023  & 0.70 & 1.35 & 0.66   \\
   & $3/2^+_2$ &  2.326  & 2.345 &  & 0.001 & $1 \times 10^{-4}$ &  0.007 
            & $1 \times 10^{-4}$  & $4 \times 10^{-4}$  & 2.31  & 0.23  & 0.04  \\
   & $5/2^+_2$ &  2.952  & 2.948 &  & 0.408 & 0.566 &  0.207 
            & $3 \times 10^{-3}$  & 0.484  & 0.84  & 1.17  &  1.41  \\
   & $1/2^+_2$  &  3.928  & 3.926 & &0.036  & 0.034 &  0.017 
            & $7 \times 10^{-6}$  & 0.035  & 1.03  & 0.97  & 1.24    \\

 \\

\hspace{-0.3cm} $\mathbf{^{35}Cl  \rightarrow ^{35}Ar} $\\
\textbf{Expt.} \cite{NNDC}\\
$A$ & $J^{\pi}$ &\hspace{2mm} $^{35}$Cl &\hspace{2mm} $^{35}$Ar & &\hspace{2mm} $^{35}$Cl &\hspace{2mm} $^{35}$Ar &\hspace{2mm} $B(GT)$ &\hspace{2mm} $B_{IS}(M1)$ &\hspace{2mm} $B_{IV}(M1)$  &\hspace{2mm} $^{35}$Cl &\hspace{2mm} $^{35}$Ar &\hspace{2mm} $R_{oc}$ \\
 \hline
\\
35 & $3/2^+_1$  &  0.0  & 0.0 &  & 0.822 \footnotemark[1] &   &   
            &   &   &   &   &      \\
   & $1/2^+_1$  &  1.220  & 1.184 & &0.085(2)  &   &   
            &   &   &   &   &      \\
   & $5/2^+_1$  &  1.763  & 1.450 & &0.003(3)  &    & 
             &   &   &   &   &      \\

\\
\hline

 \end{tabular}
	\end{table*}
\addtocounter{table}{-1}

\begin{table*}
\addtolength{\tabcolsep}{+0.3mm}
	\leavevmode
	 \centering  
	 \caption{ Continued.}\label{tab_1} 
	 \begin{tabular}{ lccccccccccccc } 
\hline
 \hline
   & &  \multicolumn{2}{c}{$E_x$} & & \multicolumn{2}{c}{$B(M1) \uparrow$ } &  &  &   &\multicolumn{2}{c}{$R_{IS}$} &   \\
  \cline{3-4}
  \cline{6-7}
  \cline{11-12}

\textbf{USDC}\\
$A$ & $J^{\pi}$ &\hspace{2mm} $^{35}$Cl &\hspace{2mm} $^{35}$Ar & &\hspace{2mm} $^{35}$Cl &\hspace{2mm} $^{35}$Ar &\hspace{2mm} $B(GT)$ &\hspace{2mm} $B_{IS}(M1)$ &\hspace{2mm} $B_{IV}(M1)$  &\hspace{2mm} $^{37}$Ar &\hspace{2mm} $^{37}$K &\hspace{2mm} $R_{oc}$ \\
 \hline
35 & $3/2^+_1$  &  0.0  & 0.0 &  & 0.878 \footnotemark[1] & 0.636 \footnotemark[1] &  0.077 
            & 0.757 \footnotemark[2] & 0.121 \footnotemark[3]& 52.65 & 27.63 &  0.04    \\
   & $1/2^+_1$  &  1.244  & 1.150 & &0.027  & 0.017 &  0.026 
            & $3 \times 10^{-4}$  & 0.022  & 1.24  & 0.78  &  0.50   \\
   & $5/2^+_1$  &  1.725  & 1.706 & &0.005  & 0.008  & 0.027
             & $9 \times 10^{-5}$  & 0.006  & 0.78  & 1.25  & 0.14    \\
   & $3/2^+_2$  & 2.688 &  2.601 & & 0.050  & 0.033   & 0.033
             & $4 \times 10^{-4}$ & 0.041 & 1.22  & 0.80  &  0.75    \\
   & $5/2^+_2$  &  3.091   & 3.059  & & 0.059  &  0.036  & 0.014
             & $7 \times 10^{-4}$ & 0.047  & 1.26  & 0.77  &  2.02    \\

\\  

\hspace{-0.3cm} $\mathbf{^{37}Ar  \rightarrow ^{37}K} $\\
\textbf{Expt.}  \cite{NNDC}\\
$A$ & $J^{\pi}$ &\hspace{2mm} $^{37}$Ar &\hspace{2mm} $^{37}$K & &\hspace{2mm} $^{37}$Ar &\hspace{2mm} $^{37}$K &\hspace{2mm} $B(GT)$ &\hspace{2mm} $B_{IS}(M1)$ &\hspace{2mm} $B_{IV}(M1)$  &\hspace{2mm} $^{37}$Ar &\hspace{2mm} $^{37}$K &\hspace{2mm} $R_{oc}$ \\
 \hline
37 & $3/2^+_1$  &  0.0  & 0.0 &  & 1.145(5) \footnotemark[1] & 0.203 \footnotemark[1]&   
            & 0.674 \footnotemark[2] & 0.471 \footnotemark[3]  & 5.91  & 0.18 &      \\
   & $1/2^+_1$  &  1.409  & 1.370 & &   &   &   
            &   &   &   &   &      \\
    & $5/2^+_1$  &  2.796  & 2.750 & &0.185(107)  & 1.96(86)  & 
             & 0.235(161)  & 0.837(303)  & 0.22(15)  & 2.34(133)  &      \\
   & $5/2^+_2$  &  3.170  & 3.239 & &0.021(6)  &    & 
             &   &   &   &   &      \\
 
\textbf{USDC}\\
37 & $3/2^+_1$  &  0.0  & 0.0 &  & 1.107 \footnotemark[1] & 0.330 \footnotemark[1] &  0.232 
            & 0.718 \footnotemark[2] & 0.388 \footnotemark[3] & 8.12 & 0.72  & 0.15    \\
   & $1/2^+_1$  &  1.439  & 1.437 & &0.016  & 0.014 & $4 \times 10^{-4}$ 
            & $1 \times 10^{-5}$ & 0.015 & 1.06 & 0.93  & 22.67     \\
   & $5/2^+_1$  &  2.729  & 2.766 & &0.273  & 0.447  & 0.504
             & 0.005  & 0.354  &  0.77 & 1.26 & 0.42   \\ 
   & $5/2^+_2$  &  3.140  & 3.165 & &0.465  & 0.550  & 0.148
             &  $9 \times 10^{-4}$ & 0.506 & 0.92 & 1.08  & 2.07    \\
   & $3/2^+_2$  &  3.641  & 3.711 &  & 0.018 & 0.013&  0.036 
            & $1 \times 10^{-4}$  & 0.015  & 1.17 & 0.84 & 0.26    \\
   & $1/2^+_2$  &  4.162  & 4.214 & &$3 \times 10^{-4}$  & $5 \times 10^{-4}$ & 0.031 
            & $6 \times 10^{-6}$ & $4 \times 10^{-4}$ & 0.76 & 1.27  & 0.01     \\
 

 \hline
 \hline
 \label{tab_1}
\end{tabular}
\footnotetext[1]{Represents $\mu$}
\footnotetext[2]{Represents $\mu_{IS}$}
\footnotetext[3]{Represents $\mu_{IV}$}
\end{table*}

\subsection{$\mathbf{^{23}Na \rightarrow^{23}Mg}$}

The $M1$ transition strengths in the $^{23}$Na and $^{23}$Mg nuclei and the corresponding GT strengths for analogous transitions between the mirror pair are well reproduced from the USDC interaction. As discussed earlier, the IS components are found to be much smaller than the IV components. For stronger $M1$ transitions i.e., from $3/2_1^+$ to $5/2_1^+$, $1/2_2^+$ and $5/2_3^+$, the GT strengths are much smaller than the $B(M1)$ values. For these cases, the $R_{IS}$ values remain approximately 1, suggesting that the IS contribution is small. This implies that, it is the constructive interference between the orbital and spin IV terms which enhances the $M1$ transition strengths. The $R_{oc}$ or $R_{ISO}$ values of these transitions also reflect the same implications. The calculated $R_{ISO}$ value is smaller than the measured value in $3/2_1^+ \rightarrow 5/2_3^+$, which is associated with large experimental uncertainty. Due to the lack of experimental data, the measured $R_{IS}$ values are available only for a few transitions. Further, except for the  $3/2_1^+ \rightarrow 5/2_1^+$ transition, only the $R_{ISO}$ values have been deduced for all transitions from the experimental data \cite{23NaMg}  due to the unavailability of $B(M1)$ strengths in $^{23}$Mg. We have calculated $R_{IS}$, $R_{oc}$, and $R_{ISO}$ parameters for all transitions corresponding to the experimental data. In the stronger transitions, the IS term has a small effect of around $ 8-10 \%$, while it modifies the $M1$ strengths by $\approx 20 \%$ in other cases. In the transitions $3/2_1^+ \rightarrow 3/2_2^+$ and $3/2_1^+ \rightarrow 5/2_2^+$, the $R_{oc}$ values suggest that the orbital parts do not interfere constructively with the spin parts in the IV $M1$ term. 
The $^{23}$Na - $^{23}$Mg nuclei are largely deformed, and the orbital contributions to different transitions in these nuclei were discussed using the deformed Nilsson's model and the particle rotor model in Ref. \cite{23NaMg}. The deformed structure and collective properties of these isotopes can be well explained from the shell model theory \cite{E2A23}. In the framework of the shell model, the energy eigenstates arise from the mixing of various configurations involving $sd$ -shell orbitals in deformed nuclei. In $^{23}$Na, the ground state emerges from the configuration $\pi (d_{5/2}^3) \otimes \nu (d_{5/2}^4)$. The major components of the wave functions of the states $5/2_1^+$, $1/2_2^+$, and $5/2_3^+$ are also the same. The transitions from the g.s. to these states are associated with large positive orbital contributions. Due to isospin symmetry, identical structures are found in the mirror partner $^{23}$Mg with proton and neutron configurations interchanged. Similar interpretations can be made for $^{23}$Mg.

\begin{figure}[h]
    \centering
    \includegraphics[scale=0.77]{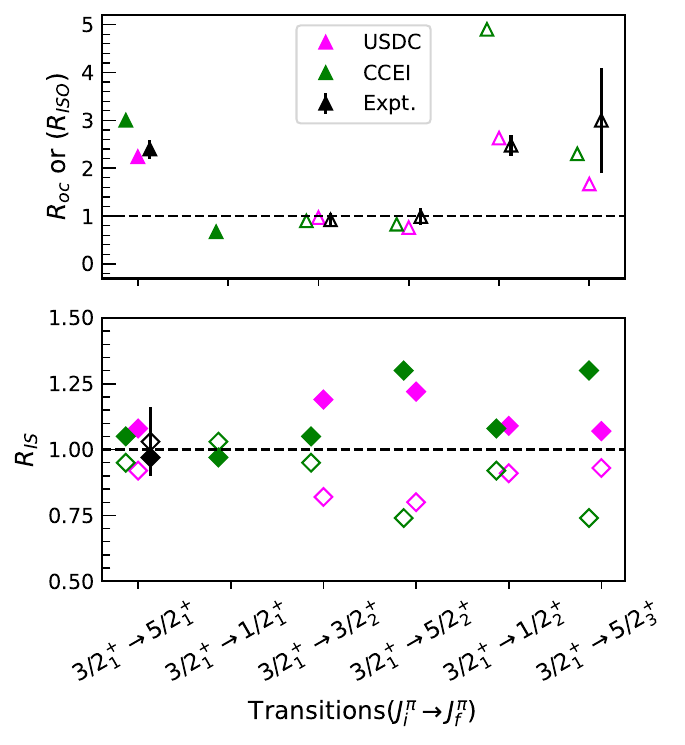}
    \caption{Upper panel: $R_{oc}$ (filled triangles) or $R_{ISO}$ (unfilled triangles) of different transitions. Lower panel: The black, magenta, and green diamonds correspond to the Expt., USDC, and CCEI data. The filled (unfilled) symbols refer to the $T_z=1/2$ ($T_z=-1/2$) nucleus in the $A=23$ pair.}
    \label{fig:A23}
\end{figure}

The \textit{ab initio} results are compared with the USDC results in Fig. \ref{fig:A23}. They lie very close to the experimental data and results obtained from the USDC interaction. Large deviation arises only in the case of $3/2_1^+ \rightarrow 1/2_2^+$. 
For this transition, a large $R_{ISO}$ value or a relatively much smaller GT transition strength is observed compared to the $B(M1)$ values in the CCEI interaction. The orbital contribution is solely responsible for such enhancement in the $M1$ transition strength since the IS contribution from CCEI interaction remains the same as that of USDC. From the experiment and USDC interaction, a very weak $M1$ strength is observed for the transition $3/2_1^+ \rightarrow 1/2_1^+$. The deduced $R_{oc}$ and $R_{IS}$ values of such weak transitions are not reliable \cite{23NaMg} and are not shown in Fig. \ref{fig:A23}. But in CCEI, the $M1$ strengths for this transition are not negligible, and its prediction is depicted in Fig. \ref{fig:A23}.

\subsection{$\mathbf{^{25}Mg \rightarrow ^{25}Al}$}
In contrast to the $A=23$ pair, weak $M1$ and GT strengths are measured for transitions between analogous states in the $A=25$ system except for the transition $5/2_1^+ \rightarrow 7/2_1^+$. The results obtained from the USDC interaction display a similar pattern. Only in the case of $5/2_1^+ \rightarrow 5/2_2^+$ transition, the calculated  GT strength is significantly smaller than the experimental data. The $R_{IS}$ remains almost 1 in $5/2_1^+ \rightarrow 7/2_1^+$, and the large $R_{oc}$ value exceeding unity implies constructive interference between the orbital and the spin IV term. 
The $A=25$ system $^{25}$Mg and $^{25}$Al are also strongly deformed \cite{DeformationOtsuka} like those $A=23$, $T_z=\pm 1/2$ nuclei. The orbital contributions to $M1$ transitions of the $A=25$ pair were been discussed in the context of deformed Nilsson's model and particle rotor model in Ref. \cite{25MgAl}. In the shell model, the g.s. in $^{25}$Mg arises from the configuration $\pi (d_{5/2}^4) \otimes \nu (d_{5/2}^5)$. The major component of the wave function configurations is $\pi (d_{5/2}^4) \otimes \nu (d_{5/2}^5)$ only in the $7/2_1^+$ state, and the $B(M1)$ strength from g.s. to this state is enhanced by the constructive interference of the orbital IV term in the $M1$ transition. Similar interpretations can be made for the mirror partner $^{25}$Al, where the corresponding configuration is $\pi (d_{5/2}^5) \otimes \nu (d_{5/2}^4)$. The results obtained from CCEI are similar to the USDC results in which the transition strengths remain strong only for the transition from $5/2_1^+$ to $7/2_1^+$. The associated $R_{oc}$ (or $R_{ISO}$) value for this transition is 3.39 (or 3.04).

As described before, the parameter $R_{oc}$ and its interpretations were based on the approximation that the IV spin term is not negligible or hindered. In weak $M1$ or GT transitions, the IV spin terms are found to be very small, and in such cases, the $R_{oc}$ values are not reliable or could not be interpreted with the usual meaning \cite{23NaMg,27AlSi,25MgAl}. Further, the measured $B(GT)$ strengths less than 0.04 units are considered to be less reliable \cite{25MgAl, Fujita_A27, Fujita2018}. Hence, from here on, the discussion and depiction through figures of the $M1$ components and $R_{oc}$ values are excluded or avoided where $B(M1)$ and $B(GT)$ transition strengths are very weak or generally smaller than 0.03 units in the text.

\subsection{$\mathbf{^{27}Al \rightarrow ^{27}Si}$}
The calculated  $M1$ and GT transition strengths in the $A=27$ pair are in good agreement with the experimental data. The corresponding $R_{oc}$ (or $R_{ISO}$) values are also well reproduced. A notable deviation in $B(M1)$ strength is observed between the calculated and experimental data for the transition $5/2_1^+ \rightarrow 5/2_3^+$ in $^{27}$Si.
The measured $B(M1)$ strength for $5/2_1^+ \rightarrow 5/2_3^+$ has a large uncertainty, and it differs nearly by $\approx$ 3 times from the respective $B(M1)$ strength of the mirror nucleus $^{27}$Al. However, in the calculated results, the difference is only  $\approx$ 1.36 times, and the $R_{oc}$ parameter for this transition is also larger than unity while the experimental value with errors is close to 1. Therefore, further measurements with more precision are required to reaffirm the orbital contributions in this transition. 
\begin{figure}[h]
    \centering
    \includegraphics[scale=0.77]{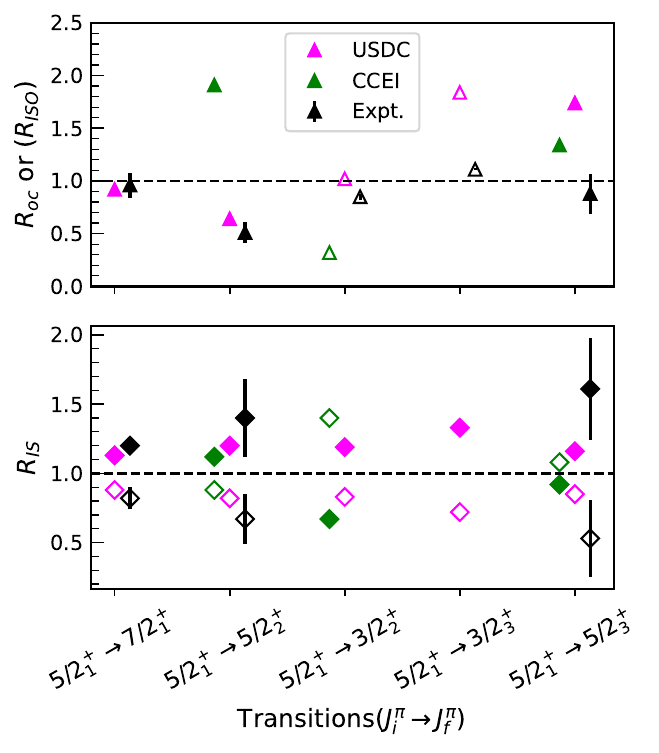}
    \caption{Same as Fig. \ref{fig:A23} but for $A=27$ pair. }
    \label{fig:A27}
\end{figure}

The mirror nuclei $^{27}$Al and $^{27}$Si are less deformed compared to the $A=23$ and 25 pair, and the energy states contain single-particle-like configurations. In such cases, the wave functions of the energy eigenstates are dominated by a single particle state or an orbital where the odd nucleon resides. The IV orbital and spin ($l \tau$ and $\sigma \tau$ ) terms are hindered for the transition from $0d$ to $0s$ orbital, while the $l \tau$ terms are facilitated between single-particle states with the same orbital angular momentum $l$ provided that $l \neq 0$ \cite{SuhonenBook}. Further, it has been observed that the $M1$ transition probabilities are enhanced for transitions between spin-orbit partners \cite{M1_32S}. Because of isospin symmetry, the structure and configurations of the $T_z= -1/2$ nucleus closely resemble those of the $T_z= 1/2$ nucleus, with only protons and neutrons configurations interchanged. So here and in the subsequent mirror pairs in the text, our discussion on the transitions will be based on the $T_z=1/2$ nucleus of the mirror pair. Analogous arguments apply to the $T_z=-1/2$ nucleus. In $^{27}$Al, the g.s. $5/2_1^+$ is formed out of a configuration $\pi (d_{5/2}^{-1})$ while the $3/2_1^+$ state is obtained from $\pi (s_{1/2}^{1})$ type configurations, i.e., when a proton in $s_{1/2}$ orbit coupled to $2_1^+$ state of $^{26}$Mg core. So, the transition between $5/2_1^+$ and $3/2_1^+$ states is strongly hindered. The $5/2_2^+$ and $5/2_3^+$ states arise from the mixing of a large number of configurations. The $B(E2)$ strengths serve as stringent probes to understand the structure of nuclear wave functions. The calculated $E2$ strength between $5/2_1^+$ and $5/2_2^+$ is 14.33 $e^2fm^4$ while it is 0.07 $e^2fm^4$ for transition from $5/2_1^+$ to $5/2_3^+$. This implies a change in the configuration of the wave function of $5/2_3^+$. This state is expected to dominate with $d_{3/2}$ orbital. A large orbital contribution is observed for the transition between $5/2_1^+$ and $5/2_3^+$ or between the spin-orbit partners. The  $3/2_3^+$ and $7/2_2^+$ states are also of such nature, and the $E2$ strengths from $5/2_1^+$ to these states are 0.04 $e^2fm^4$ and 1.75 $e^2fm^4$ respectively. The $R_{oc}$ value for the transition from g.s. to these states remains larger than 1. 

Within CCEI, very weak transition strengths are predicted for transitions other than $5/2_1^+$ to $5/2_2^+$, $3/2_2^+$, and $5/2_3^+$. The CCEI results are compared with USDC results in Fig. \ref{fig:A27}, wherever the transition strengths are not negligible. While the IS contributions are approximately the same or exhibit small differences compared to those found in USDC, the predicted $R_{oc}$ (or $R_{ISO}$) parameters suffer deviations from those obtained with the USDC interaction.

\subsection{$\mathbf{^{29}Si \rightarrow ^{29}P}$}
The GT transition strengths have not yet been measured between the $^{29}$Si - $^{29}$P mirror pair. We have calculated the $M1$ transition strengths from g.s. to several excited states and the $B(GT)$ values for corresponding analogous states in the $A=29$ pair. The parameter $R_{IS}$ is a signature of IS contribution to the $M1$ transition. Though the $B_{IS}(M1)$ values are small, the $R_{IS}$ varies in the range 0.7-1.3, implying that the $M1$ transition strengths can be modified by $\pm 30\%$ in the mirror pair due to constructive or destructive contributions of the IS term. 
\begin{figure}
    \centering
    \includegraphics[scale=0.77]{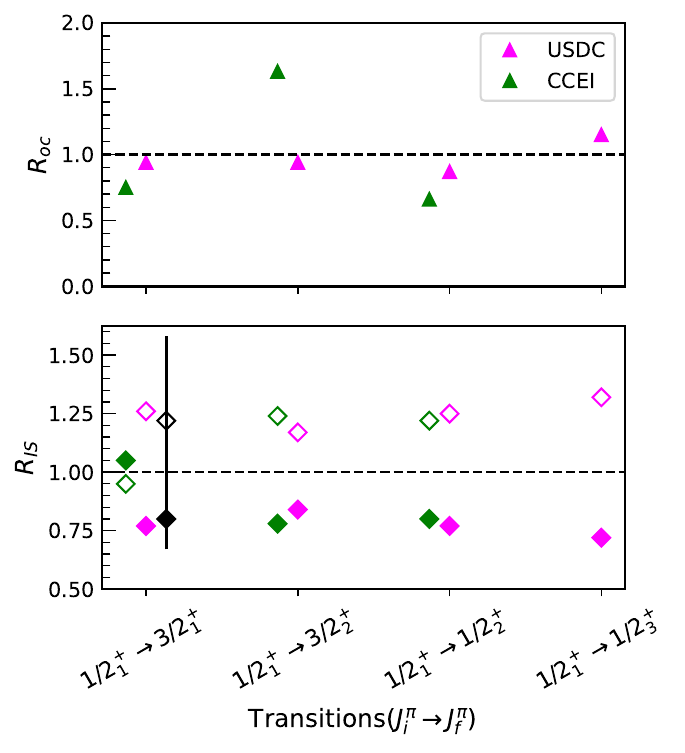}
    \caption{Same as Fig. \ref{fig:A23} but for $A=29$ pair. }
    \label{fig:A29}
\end{figure}
The g.s. $1/2_1^+$ of $^{29}$Si is described by the $\pi (d_{5/2}^6) \otimes \nu (d_{5/2}^6 s_{1/2}^1)$ configuration. The major configuration of $3/2_1^+$ and $3/2_3^+$ states is $\pi (d_{5/2}^6) \otimes \nu (d_{5/2}^6 d_{3/2}^1)$ or is dominated by $d_{3/2}$ orbital. So, weak transition strengths are observed for the transition from g.s. to $3/2_1^+$ and $3/2_3^+$. If we look at the $E2$ transition probabilities, the $B(E2;1/2_1^+ \rightarrow 3/2_3^+)$=0.2 $e^2fm^4$ and $B(E2;1/2_1^+ \rightarrow 3/2_1^+)$=58.7 $e^2fm^4$, respectively. It implies that the $3/2_3^+$ has a completely different configuration than $1/2_1^+$ or characterized by $d_{3/2}$ orbital, and the transition from $1/2_1^+$ to $3/2_3^+$ is strongly hindered. The major component of the wave function of $1/2_2^+$ is the same as the $1/2_1^+$ state, while $3/2_2^+$ and $1/2_3^+$ appear from large configuration mixing. The $R_{oc}$ parameters for the transitions from g.s. to these states are close to 1, suggesting that the orbital term has negligible or no contributions in these transitions. This is evident since the orbital IV term ($l \tau$) vanishes between $0s$ states.

The CCEI results are compared with those obtained from USDC in Fig. \ref{fig:A29}. In addition to $1/2_1^+ \rightarrow 3/2_3^+$, the CCEI also predicts weak transition strengths for $1/2_1^+ \rightarrow 1/2_3^+$ and are not shown in Fig. \ref{fig:A29}. The CCEI interaction predicts a large orbital contribution to the IV term of $M1$ operator for the transition $1/2_1^+ \rightarrow 3/2_2^+$ while it is $\approx 1$ in the USDC interaction. For other transitions, the calculated $R_{oc}$ values from CCEI are close to USDC results.

\subsection{$\mathbf{^{31}P \rightarrow ^{31}S}$}
\begin{figure}[h]
    \centering
    \includegraphics[scale=0.77]{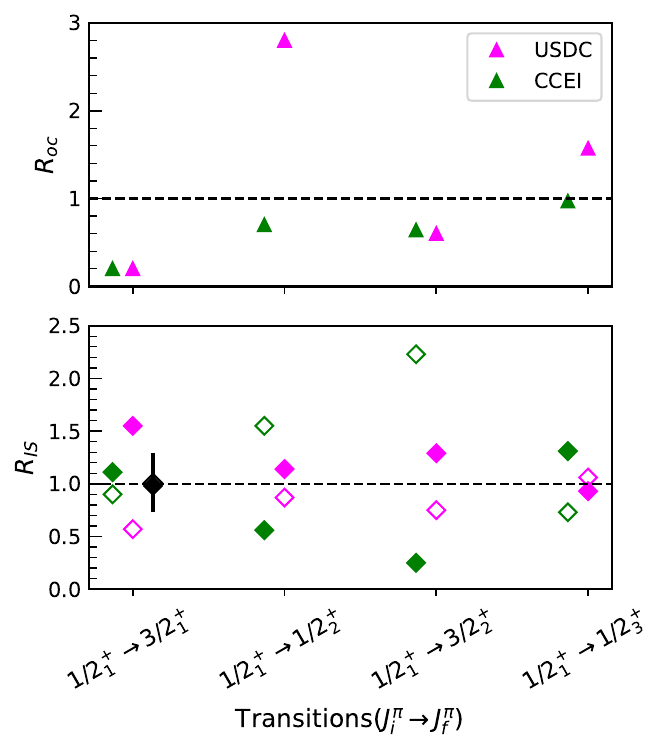}
    \caption{Same as Fig. \ref{fig:A23} but for $A=31$ pair. }
    \label{fig:A31}
\end{figure}

Few experimental data are available on the $M1$ transition strengths in the $A=31$ mirror pair. We have reported the $B(M1)$ strengths from g.s. to various excited states as well as the $B(GT)$ values for the analogous transitions. The low-energy excited states of the mirror nuclei $^{31}$P - $^{31}$S exhibit patterns similar to those of the $A=29$ pair. In $^{31}$P, the g.s. is obtained from a $\pi (s_{1/2})$ type configuration. 
But the wave functions of $1/2_2^+$ and $1/2_3^+$ states are rather mixed and dominated by the mixture of $\pi (d_{5/2}^6 s_{1/2}^1) \otimes \nu (d_{5/2}^6 s_{1/2}^2)$ and $\pi (d_{5/2}^6 s_{1/2}^1) \otimes \nu (d_{5/2}^6 s_{1/2}^1 d_{3/2}^1)$ configurations. The $M1$ transitions from $1/2_1^+$ to $1/2_2^+$ and $1/2_3^+$ states remain strong. However, the orbital contributions in these transitions are different than those of the $A=29$, $T_z=\pm 1/2$ pair. Unlike the $A=29$ pair, these transitions are associated with large orbital contributions, which implies that they may have large deformations. Again, the major configurations of $3/2_1^+$ and $3/2_3^+$ states are dominated with $d_{3/2}$ orbital while the $3/2_2^+$ state is obtained from large configurations mixing. So, the transition strengths from g.s. to $3/2_1^+$ and $3/2_3^+$ are small or negligible. The IS and orbital contributions to the transitions in the $A=31$ mirror pair calculated from CCEI  are displayed in Fig. \ref{fig:A31}. The transition strengths from g.s. to $3/2_3^+$ are very weak and are not shown. In CCEI, the $B(M1)$ strengths are not as strong as those observed in USDC in $1/2_1^+ \rightarrow 1/2_2^+$, and the orbital IV term interferes destructively with the spin IV term. Also, the IS contributions to mirror pairs in the transition $1/2_1^+ \rightarrow 3/2_2^+$ calculated from CCEI differ from those obtained with USDC.


\subsection{$\mathbf{^{33}S \rightarrow ^{33}Cl}$}
The calculated $M1$ transition strengths for $A=33$ pair are compared with the corresponding experimental data in Table \ref{tab_1}. We have reported the $B(GT)$ strengths of the analogous transitions between $^{33}$S - $^{33}$Cl nuclei for which experimental data are not available. In $^{33}$S, the g.s. $3/2_1^+$ comes out from a configuration $\pi (d_{5/2}^6 s_{1/2}^2 ) \otimes \nu (d_{5/2}^6 s_{1/2}^2 d_{3/2}^1)$. The $1/2_1^+$ state arises from a configuration $\pi (d_{5/2}^6 s_{1/2}^2 ) \otimes \nu (d_{5/2}^6 s_{1/2}^1 d_{3/2}^2)$ or characterized by $\nu s_{1/2}$ orbital. The $3/2_2^+$ state is also found to exhibit such characters. So, the transitions from $3/2_1^+$ to $1/2_1^+$ and $3/2_2^+$ are strongly hindered. The transition strengths from g.s. to $1/2_2^+$ are also weak. The dominant wave function configuration of the $5/2_2^+$ state $\pi (d_{5/2}^6 s_{1/2}^1 d_{3/2}^1 ) \otimes \nu (d_{5/2}^6 s_{1/2}^1 d_{3/2}^2)$ is a mixture of $sd$ orbitals. The g.s. and $5/2_2^+$ state are connected with a strong $M1$ transition, and a positive orbital contribution is observed. But the $M1$ strength is weak in the transition $3/2_1^+ \rightarrow 5/2_1^+$. Because the major constituent of the $5/2_1^+$ state is  $\pi (d_{5/2}^6 s_{1/2}^2 ) \otimes \nu (d_{5/2}^6 s_{1/2}^1 d_{3/2}^2)$, this state can be interpreted as a neutron occupying in the $s_{1/2}$ orbital coupled to the $2_1^+$ state of core $^{32}$S. Transition strengths from CCEI interaction for $3/2_1^+ \rightarrow 1/2_1^+$ and $3/2_1^+ \rightarrow 3/2_2^+$ are also observed to be very weak. The CCEI results are similar to USDC results for the transition $3/2_1^+ \rightarrow 5/2_1^+$, but they diverge for the transition $3/2_1^+ \rightarrow 5/2_2^+$. It also predicts a very small $B(GT)$ strength of less than 0.02 units for the transition $3/2_1^+ \rightarrow 1/2_2^+$.

\vspace{0.3cm}
\subsection{$\mathbf{^{35}Cl \rightarrow ^{35}Ar}$}
The experimental data on $M1$ and GT transition strengths are limited in the $^{35}$Cl - $^{35}$Ar pair, particularly in $^{35}$Ar. We have reported several states and provided estimates of the $B(M1)$ and $B(GT)$ strengths for various transitions. The calculated transition strengths are found to be either weak or of moderate magnitude, generally $\approx 0.05$ units. Like other mirror pairs, transition strengths larger than 0.1 units are not observed. The orbital IV term interferes constructively with the spin IV term only in the transition from g.s. to $5/2_2^+$. The g.s. $3/2_1^+$ has a configuration $\pi (d_{5/2}^6 s_{1/2}^2 d_{3/2}^1 ) \otimes \nu (d_{5/2}^6 s_{1/2}^2 d_{3/2}^2)$ in $^{35}$Cl. The main components of the wave function of $5/2_1^+$ state are $\pi (d_{5/2}^6 s_{1/2}^1 d_{3/2}^2 ) \otimes \nu (d_{5/2}^6 s_{1/2}^2 d_{3/2}^2)$ and $\pi (d_{5/2}^6 s_{1/2}^2 d_{3/2}^1 ) \otimes \nu (d_{5/2}^6 s_{1/2}^2 d_{3/2}^2)$ while those in the $5/2_2^+$ state are $\pi (d_{5/2}^6 s_{1/2}^1 d_{3/2}^2 ) \otimes \nu (d_{5/2}^6 s_{1/2}^2 d_{3/2}^2)$ and $\pi (d_{5/2}^5 s_{1/2}^2 d_{3/2}^2 ) \otimes \nu (d_{5/2}^6 s_{1/2}^2 d_{3/2}^2)$. This suggests that the $d_{5/2}$ orbital character present in the $5/2_2^+$ state is the possible cause of a large $R_{oc}$ parameter in the transition from g.s. to this state. Within the CCEI framework, the $M1$ and GT strengths for transitions involving g.s. and $1/2_1^+$, $3/2_2^+$ are very small, typically $\leq 0.01$ units. In the transition from g.s. to $5/2_1^+$, the $B(M1)$ strengths are very weak, while the $B(GT)$ strength is negligible in the transition $3/2_1^+ \rightarrow 5/2_2^+$.

\subsection{$\mathbf{^{37}Ar \rightarrow ^{37}K}$}
\begin{figure}[h]
    \centering
    \includegraphics[scale=0.77]{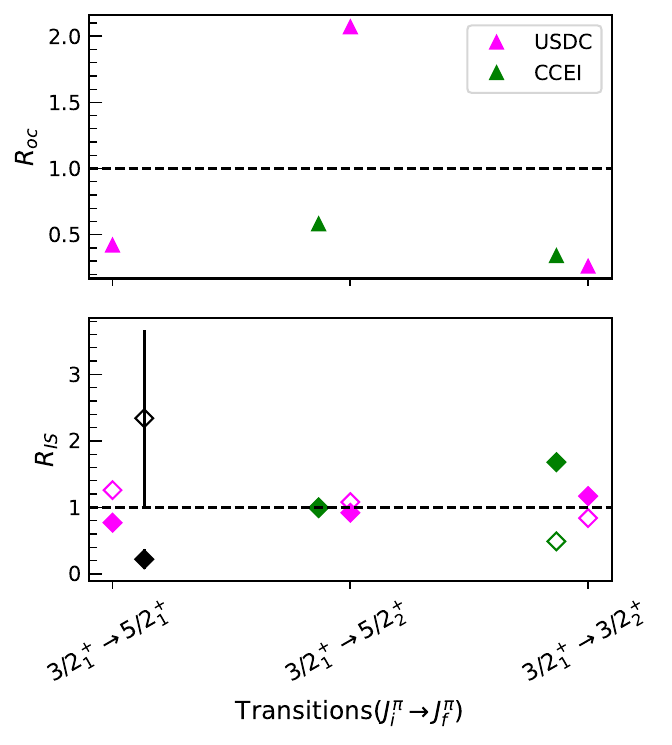}
    \caption{Same as Fig. \ref{fig:A23} but for $A=37$ pair. }
    \label{fig:A37}
\end{figure}

\begin{figure*}
    \centering
    \includegraphics[scale=0.75]{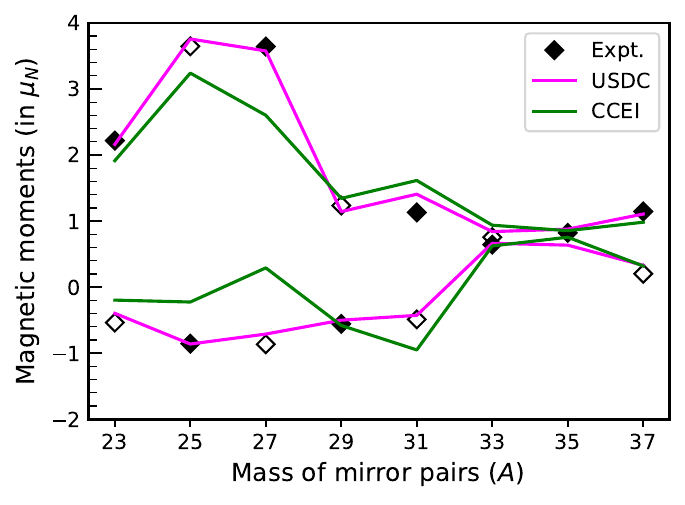}
    \includegraphics[scale=0.75]{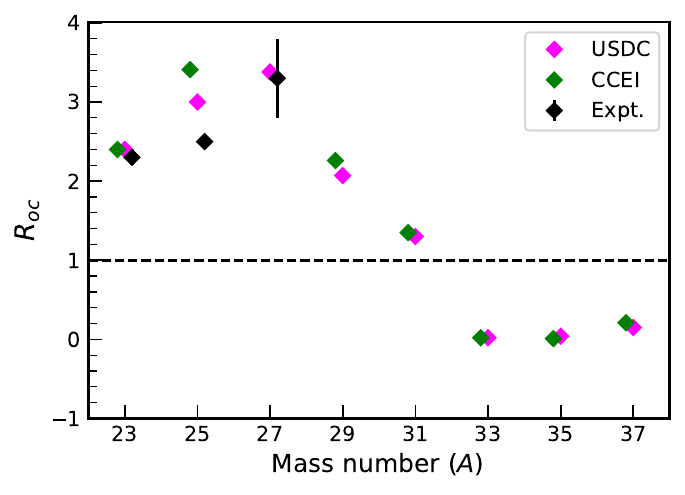}
    
    \caption{Left panel: The filled (unfilled) black symbols refer to the experimental data for g.s. magnetic moments of the $T_z=1/2$ ($T_z=-1/2$) nucleus, while the color lines are their corresponding calculated results.} Right panel: $R_{oc}$ parameter in the g.s. magnetic moments of the mirror pairs.
    \label{fig:Roc_gs_mu}
\end{figure*}
For the mirror pair $^{37}$Ar - $^{37}$K, limited experimental data are available, and they are associated with large uncertainty. The $M1$ and GT transition strengths between g.s. and lower excited states are calculated from the USDC interaction. The $M1$ and GT strengths remain strong for the transitions $3/2_1^+ \rightarrow 5/2_1^+$ and $3/2_1^+ \rightarrow 5/2_2^+$. Both $3/2_1^+$ and $5/2_1^+$ arise from a configuration $\pi (d_{5/2}^6 s_{1/2}^2 d_{3/2}^2 ) \otimes \nu (d_{5/2}^6 s_{1/2}^2 d_{3/2}^3)$ in $^{37}$Ar. But the major components of wave function configurations of $5/2_2^+$ state are of $\nu (d_{5/2}^{-1})$ type. The transition from g.s. to $5/2_2^+$ or between the spin-orbit partners is accompanied by constructive interference of the orbital term in the $M1$ transition. The USDC results are compared with the calculated results obtained from CCEI in Fig. \ref{fig:A37}. In CCEI, the transition from $3/2_1^+$ to $5/2_1^+$ is associated with negligible $M1$ strength, and its results are not shown in the figure. The CCEI could not reproduce the $R_{oc}$ parameter in the transition $3/2_1^+ \rightarrow 5/2_2^+$.

\subsection*{Magnetic moments:}

In this section, the ground state magnetic moments and their respective components are analyzed for the mirror pairs. The calculated g.s. magnetic moments of the $T_z=\pm 1/2$ pair, along with the experimental data, are shown in the left panel of Fig. \ref{fig:Roc_gs_mu}. We can see that all available measured data are well reproduced from the USDC interaction. The CCEI results also lie close to the experimental data or USDC results. The orbital contribution to the ground state magnetic moments or $R_{oc}$ parameters are shown on the right panel of Fig. \ref{fig:Roc_gs_mu}. Both USDC and CCEI make similar predictions for $R_{oc}$ values in all mirror pairs except the $A=27$ pair. The g.s. magnetic moments obtained with CCEI for the $A=27$ pair exhibit larger differences from the corresponding experimental data compared to other pairs (see left panel of  Fig. \ref{fig:Roc_gs_mu}). The CCEI also fails to reproduce the GT strength between ground states in the $^{27}$Al - $^{27}$Si pair, which results in a large $R_{oc}$ value ($\approx 7$) and is out of the scale of Fig. \ref{fig:Roc_gs_mu}. The parameter $R_{oc}$ has large values ($R_{oc} >1$) in $A=23$ to $31$ but is reduced significantly or lies near zero in the mirror pairs with mass $A=33$ to $37$. 

Large asymmetry is observed in the g.s. magnetic moments of the $A=23$ to $27$ pairs, or the magnetic moments of these $T_z=\pm 1/2$ pairs have huge differences. This difference is decreased in $A=29$ and 31 and further reduced significantly in the mirror nuclei $A=33$ to $37$. Accordingly, the $R_{oc}$ parameters are large within the $A=23$ to $27$ mass range, decrease at $A=29$ and $31$, but are still larger than 1. Then, they reduced to nearly zero in $A=33$ to $37$. Actually, when the magnetic moments of the mirror pairs exhibit substantial differences, the corresponding IV part or $\mu_{IV}$ components are larger, resulting in an increased $R_{oc}$ parameter [see Eq. \eqref{eq15}]. From Table \ref{tab_1}, it can be observed that the corresponding IS components are smaller than the IV components in these nuclei. However, in $A=33$ to 37 pairs, the magnetic moments of $T_z=1/2$ and $T_z=-1/2$ nuclei have similar values, the associated IS terms are large while the IV components are small, leading to a reduction in the $R_{oc}$ parameter.

As discussed before, the ground state wave functions in the $A=33$ to $37$ nuclei show single-particle structures. The calculated magnetic moments also do not differ much from the corresponding Schmidt value or single-particle magnetic moments. The IV orbital term destructively interferes in the $M1$ transition between the ground states of these mirror pairs. 
\begin{figure}[h]
    \centering
    \includegraphics[scale=0.75]{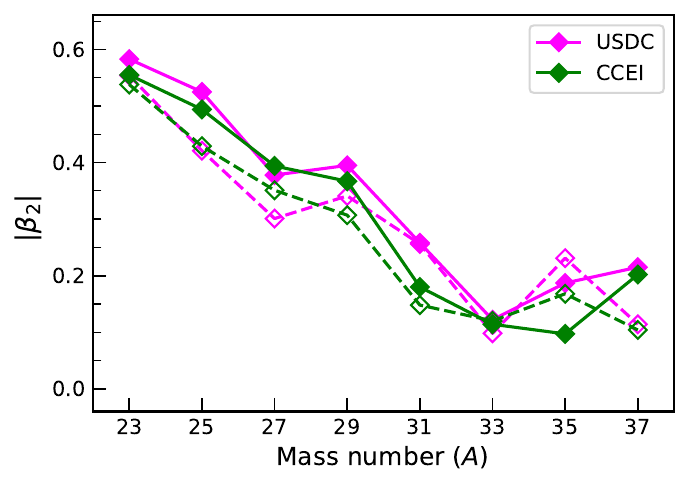}
    \caption{ The filled (unfilled) symbols are g.s. quadrupole deformation in the $T_z=1/2$ ($T_z=-1/2$) nucleus of the mirror pairs joined by a solid (dashed) line.}
    \label{fig:Beta2_gs}
\end{figure}
The $1/2_1^+$ ground states in $A=29$ and $31$ pair have $\pi (s_{1/2})$ or $\nu (s_{1/2})$ like characters. Since the orbital contributions vanish for transitions between $s$ orbits, their $R_{oc}$ values should lie close to 1, which is seen in the case of the $A=31$ pair.  However, a higher $R_{oc}$ value in the $A=29$ mirror pair suggests possible ground state deformation. To visualize the deformation in the ground states and its impact on the corresponding $R_{oc}$ values, we have plotted the quadrupole deformation parameter ( $\vert \beta_2 \vert$ ) against the mass number for the $T=1/2$ pairs in Fig. \ref{fig:Beta2_gs}. The deformation parameter is calculated using the formula \cite{Beta2}
\begin{equation}
   \vert \beta_2 \vert = \frac{4\pi}{3ZR_0^2} \times \left| \frac{\sqrt{B(E2;J_f \rightarrow J_i)}}{\langle J_f K 2 0 | J_i K \rangle} \right|,
\end{equation}
where $K$ is the projection of nuclear spin on the symmetry axis, $R_0=1.2A^{1/3}$, $J_i$ is the g.s., and $J_f$ corresponds to the excited state of the g.s. band. It can be noted that $J_f$ may or may not necessarily be the same as the first excited state and is identified from the g.s. band structures of the mirror pairs \cite{NNDC, gsBand29, gsBand31, gsBand33, gsBand35, gsBand37}. The $\vert \beta_2 \vert$ values are obtained with the standard assumption that $K=J_i$. Both USDC and CCEI make similar predictions for deformation parameters in all mirror pairs. From Fig. \ref{fig:Beta2_gs}, it can be observed that the mirror nuclei $A=23$, 25 are largely deformed, followed by $A=27$, 29 pairs, and are associated with large positive orbital contributions. The deformation is reduced in $A=31$ where $R_{oc} \approx 1$. The ground states of the $A=33$ to 37 mirror pairs are characterized by small deformation and are dominated by single-particle structures. The orbital terms ($l \tau$) have a destructive effect on the g.s. magnetic moments in these nuclei, and the $R_{oc}$ values lie close to zero.


\section{Summary} \label{Sec4}
The analogous $M1$ and GT transitions of the odd mass $T=1/2$ mirror pairs were studied using isospin non-conserving phenomenological and \textit{ab initio} effective interactions within $sd$- model space. The IS, IV components of the $M1$ operator, and the IS and orbital contributions to $M1$ transitions, are discussed in detail for various $M1$ transitions in each mirror pair. The USDC interaction well reproduces the available experimental data and is used for predictions between low-lying states where the experimental data are unavailable. Along with USDC results, the \textit{ab initio} CCEI predictions are also shown.  In several cases, the CCEI provides predictions similar to those obtained using the USDC interaction. In the mirror pairs, the IS contributions are relatively small, accounting for only about  $\approx 10 \%$ to $20 \%$ change in the $M1$ transitions. In the strong $M1$ transitions, the IV component, or particularly the spin IV part, is dominant where the orbital IV term interferes constructively or destructively depending upon the configurations of initial and final states. The g.s. magnetic moments of mirror nuclei are well explained from both USDC and CCEI interactions. Their respective IS, IV components and orbital contributions are studied and analyzed. The deformed ground states are found to exhibit large orbital contributions. 

\section{Acknowledgment} \label{Sec5}
S.S. would like to thank UGC (University Grant Commission), India, for providing financial support for his Ph.D. thesis work. P.C.S. acknowledges a research grant from SERB (India), CRG/2022/005167.



\end{document}